\documentclass[showpacs,preprint,aps]{revtex4}

\usepackage{amssymb}
\usepackage{graphicx}
\usepackage{bm}
\usepackage{footnote}

\begin{document}

\title{Magnetostatic Spin Waves and %
Magnetic-Wave Chaos in Ferromagnetic Films. \\ %
II. Numerical Simulations of Non-Linear Waves}

\author{Yu.E. Kuzovlev, Yu.V. Medvedev, and N.I. Mezin}
\email{kuzovlev@fti.dn.ua}

\affiliation{A.A.Galkin Physics and Technology Institute of NASU, ul.
R.Luxemburg 72, 83114 Donetsk, Ukraine}


\begin{abstract}
A method and some results of numeric simulations of %
magnetostatic spin waves in ferromagnetic films are %
exponded, in comparison with the theory earlier presented in %
arXiv preprint 1204.0200. In particular, roles of films %
finiteness (edges) and defects in formation of linear %
and non-linear magnetostatic wave patterns, excitation %
and evolution of two-dimensional solitons, and chaotic %
non-linear ferromagnetic resonance are considered.
\end{abstract}


\maketitle


\section*{Introduction}

This preprint represents first, preparatory, stage of %
 numerical investigation of magnetostatic spin waves' chaos %
in ferrite films realized between  2001 and 2003 %
under particular support of Multimanetic Solutions Ltd. %
Its main contents will be submitted in separate preprint, %
as Part III of the manuscript whose Part I was devoted %
to theory of magnetostatic waves and already presented by %
our arXiv preprint\, 1204.0200\,. %

Our primary purposes here, in Part II, %
and in the next Part III were:\, %
(i) to test numeric algorithms based on spatial %
discretization of film's volume;\, %
(ii) to estimate an extent to what %
the theory developed for idealized infinite films %
is applicable %
to real finite-size (and may be defective) films;\, %
(iii) to examine theoretical concepts, - %
e.g. ``quasi-local magnetic energy density'' (see below), - %
which have no unambiguous theoretical definition %
(because of strong non-locality od dipole interactions %
or by other reasons) but can be useful for %
description of non-linear magnetic-wave phenomena;\, %
(iv) to visually indicate mechanisms and forms %
of magnetostatic-wave chaos,\, %
(v) to see what of them %
are most appropriate for practical use,\, %
and numerically investigare possibilities of %
control and synchronization of this chaos. %


\section*{6. NUMERICAL METHODS AND NUMERICAL %
SIMULATIONS OF EXTERNALLY DRIVEN FILMS}


 The modern theory of nonlinear wave processes in ferromagnets
and ferrites is not developed to an extent sufficient for fruitful
applications to so complex phenomenon as magnetic chaos. By this
reason, numerical simulations must be in anyway useful, since they

(i) serve as powerful ``microscope'' to watch for magnetization dynamics at
its natural temporal scales, from the one tenth part of nanosecond up to
tens microseconds, and in this way allow to

(ii) verify existing theoretical models, find prompts for improving
analytical theory, and

(iii) obtain concrete practically acceptable estimates, conclusions and
predictions.

Although many aspects on nonlinear wave dynamics may be modelled in terms of
popular NSE (nonlinear Shr\"{o}dinger equation, see Sec.5), it does not
approach for above purposes, because it contains no natural amplitude limit
for waves and solitons. Indeed, in Eq.11 itself and in its solitonic
solutions (5.25) and (5.26) the amplitudes, \,$\, |\Psi |\,$\,  and \,$\, A\,\,$\, , may be
arbitrary large, while physically, according to relations (5.9), they can
not exceed a level of order of unit. This means that at certain critical
time moments of chaotic dynamics other higher-order non-linearities play
significant role. Besides, NSE neglects odd powers of non-linearity and
related parametric processes.

Therefore, numerical simulation algorithms should be based on complete
Landau-Lifshitz-Gilbert equation (2.1). At this approach, trivial normalization of
spin length to unit ensures taking into account all orders of
non-linearity.

6.1. NUMERICAL ALGORITHM.

To solve Eq.2.1, the classical third-order Runge-Kutta algorithm was
used, in its adaptive version with time step being automatically
chosen as large as possible at fixed precision. Nevertheless, running
of Eq.2.1 appears not a fast procedure. The matter is that non-local
dipole interaction should be calculated by means of Fourier
transform, which even in its most fast form takes much greater
time than, say, calculation of gradients and Laplasians.

Moreover, if the transform was located within the sample volume %
then the non-local interaction would be represented by a %
function of two spatial points, \,$\, r_{1}\,\,$\, and \,$\, %
r_{2}\,\,$\, instead of only their difference \ \,$\, r_{1}-r_{2}\,\,$\,,  %
and Fourier transform could not be applied. Therefore, %
when discretizing the sample volume %
into \,$\, N_{x}\times \,$\,  \,$\, N_{y}\times \,$\,  \,$\, N_{z}\,$\, %
cells (``spins''),  it is necessary to consider at least %
\,$\, N_{x}^\prime\times N_{y}^\prime\times N_{z}^\prime\,$\, %
 cells, - with %
\,$\, N_{x,y,z}^\prime\,\geq\, 2N_{x,y,z}-1\,$\,, - %
in order to correctly represent the dipole interaction kernel %
in the \,$\,(r{1}-r_{2})\,$-space %
(fortunately, \,$\, N_{x,y,z}^\prime = 2N_{x,y,z}-1\,$\, %
or \,$\, N_{x,y,z}^\prime = 2N_{x,y,z}\,$\, are %
always sufficient numbers). %
The award %
for this complication is that all the finite-size and boundary effects are
taken into account.

The algorithm was tested by modeling magnetization of small
ferromagnetic slabs, with sizes of order of tens \,$\, r_{0}\,\,$\,
(exchange interaction radius). It was possible to observe (i)
formation of magnetic vortices and domains at sufficiently weak
field, (ii) their death at moderate fields, with residual
demagnetization at former domain boundaries, (iii) almost uniform
magnetization at strong fields with significant demagnetization at
sample boundaries only, (iv) hysteretic effects, characteristic
hysteresis curves on \,$\, H-B\,$\, -plane, re-magnetization chaos
and noise at periodically varying field %
(chaotic hysteresis).

6.2. CONCRETIZATION OF NUMERICAL\ TASKS.

We will be most interested in comparatively large-scale phenomena in YIG
films whose typical size is about \,$\, 7\,\,$\, mm\,  \,$\, \times 2\,\,$\, mm\,
\,$\, \times 10\,\,$\, micron. Besides, we want to apply such kind of film
magnetization which ensures maximum MW frequencies at minimum value of
magnetizing external field \,$\, H_{0}\,\,$\, . This requirement is satisfied if (i)
film is tangentially magnetized and (ii) the surface magnetostatic waves
(MSW) are explored which propagate perpendicularly to static magnetization
(see Secs.3-4). Practically suitable wave length, \,$\, \lambda \,\,$\, \ , of these
MSW lies in the interval \,$\, 0.1\div 1\,\,$\, mm , that is \,$\, \lambda /D\gtrsim 10\,$\,  (\,$\, %
D\,\,$\, \ is film thickness), while suitable cross sizes of MSW inductors
(antennas) are comparable with \,$\, D\,\,$\, .

6.3. ASSUMPTIONS AND\ SIMPLIFICATIONS.

If we took the \,$\, r_{0}\,\,$\, (exchange radius \,$\, \sim 5\cdot 10^{-6}\,\,$\, cm)
\,  be the scale of spatial discretization, then (at above mentioned
film sizes) \,$\, N_{x}\times \,$\,  \,$\, N_{y}\times \,$\,  \,$\, %
N_{z}\,$\,  would be \,$\, \sim 10^{12}\,\,$\,.
Hence, literal simulation of real samples is impossible.

The natural possibility to simplify numerical problems arises from the fact
that surface MSW with \,$\, \lambda /D\gtrsim 10\,$\,  are almost %
uniform across film's thickness (i.e. in \,$\, z\,$\, -direction). %
Therefore, we can use the averaging over %
thickness. Then the latter becomes basic spatial scale, and we come to
two-dimensional discretization lattice, with \,$\, N_{z}=1\,\,$\, \ and \,$\, N_{x}\times \,$\,
\,$\, N_{y}\,\,$\, determined by the film length to thickness and wide to thickness
ratios.

In this approach, we inevitably neglect \,  bulk MSW modes with \,$\, %
N>0\,\ \,$\, which are essentially non-uniform %
in respect to thickness (see  Part I, Sec.4). %
These non-uniform bulk waves are not directly excitable by thick
antennae, but they can be generated from surface MSW by means of nonlinear
fourth-order G-P-processes or third-order P-processes (Sec.5) and thus
influence their dynamics. Nevertheless, %
our formally crucial simplification has physical grounds as follow.

In fact, there are similar fourth-order interactions between surface MSW
themselves. We can suppose that just these interactions are dominating in
surface MSW dynamics because realize in resonant way, while interaction %
with most of non-uniform bulk modes is far from resonances. Indeed, their
frequencies lie below uniform precession frequency, %
\,$\, \omega _{u}\,\,$\, , while %
frequencies of surface MSW are higher than \,$\, \omega _{u}\,\,$\,. %
If comparing the %
dispersion law (3.38) for DE-waves and Eqs.4.28 and 4.29 %
for dispersion of bulk modes, one can verify that their frequencies %
are close at

\begin{equation}
q_{N\pm }\approx \frac{\pi N}{D}\sim \ \frac{2\pi }{r_{0}}\sqrt{\frac{2\pi D%
}{(H_{0}+2\pi )\lambda }}\sim 1.5\cdot 10^{5}\,\text{cm}^{-1}\,,\,\ \
\,N\sim 40\,\,\,,
\end{equation}
only. In reality, so short-wave modes must be damped certainly
stronger than long surface waves. At the same time, the latter
effectively influence one upon another around %
(in \,$\,k\,$-space) equi-frequency curves %
(shown in Fig.4a, see Sec.4). By these reasons, we expect that
fourth-order interaction with non-uniform modes is relatively
insignificant. The same can be argued in respect to third-order
parametric 
by surface waves (all the more, at \,$\, %
\,\omega _{u}/2<\,$\,  \,$\, \omega _{1}\,\,$\, , i.e. at\, \ \,$\, H_{0}\gtrsim 4\pi
/3\, \,$\, , this process is forbidden at all).

At present this is (i.e. was ten years ago) unavoidable simplification, since at\, \,$\, %
N_{z}\gtrsim 2N\,\,$\, \ , that is at %
\,$\, N_{z}\,$\,  \,$\, \gtrsim 80\,$\,, again numerical
simulation would be unrealistic even with Pentium-IV in our order.
Rigorous analysis of role of non-uniform (multi-layered) MW in nonlinear
long-wave dynamics is the interesting task for future. %
Now, we yet are forced to %
deal with numerical model rather than with literal numerical simulation.

6.4. NUMERICAL STRATEGY.

After the simplification, film thickness takes the role of length unit.
Moreover, the size of discretization lattice, in units of \,$\, D\,$\,  , can be
chosen\, \ \,$\, n\times n\times 1\,\,$\, \,  , which allows to simulate
the film area \,$\, \,nN_{x}D\times \,$\,  \,$\, nN_{y}D\,\,$\, . The permissible additional
roughening, \,$\, n\,\,$\, \ , depends on characteristic minimum wave length in a
situation under analysis. To speed up calculations, we may choose a greater \,$\, %
n\,$\,  but then lower it if necessary.

In fact, the values \,$\, 1\leq n\leq 8\,\,$\,, %
\ \,$\, 20\leq N_{y}=\leq 170\,\,$\,, and %
\,$\, 70\leq N_{x}\leq 256\,$\,  %
were used\, [\,ten years ago, while %
now it is possible to take greater %
\,$\,N_x\,,\, N_y\,$\, and simultaneously %
\,$\,N_z>1\,$\,, i.e. fractional $\,\,n\,$\,]. %
Of course, the dipole interaction of discrete cells (effective spins)
was calculated with taking into account their shape as determined by
\,$\, n\,\,$\,. A choice of definite %
``magic'' \,$\, N_{x}\,$\,  and \,$\, N_{y}\,$\,  numbers ensured %
most fast FFT on a \,$\,N_{x}^\prime\times %
N_{y}^\prime\,\,$\,  %
lattice (with \,$\,N_{x,y}^\prime\geq %
 2N_{x,y}-1\,$). %
Nevertheless, typically %
from 1 to 5 real time seconds were elapsed per one period of spin
precession, \,$\, \sim 0.3\,$\, \,  ns\,, since several FFT's and inverse FFT's
should be performed at each time step.

At given uniform magnetizing field \,$\, H_{0}\,\,$\, , firstly static magnetization
pattern was calculated and conserved in memory, then serving as initial
ground state for wave and soliton structures caused by time-varying currents.

The anisotropy is what can be simply taken into account with no numerical
problems. But it evolves several parameters at once. At present stage, we
want to obtain numerical ``reference point data'' with minimum amount of
free parameters and therefore intentionally omit anisotropy.

6.5. NUMERICAL FRICTION.

It is well known that the time discretization when numerically solving
differential dynamic equations inevitably results in more or less effective
friction (or may be negative one). In our case this artifact also takes
place leading to energy relaxation even if the friction coefficient is put
on be zero, \,$\, \gamma =0\,\,$\,. Interestingly, this excess numerical relaxation
excellently obeys exponential law and therefore works as increase of \,$\, \gamma
\,\,$\, , \,$\, \gamma \rightarrow \,$\,  \,$\, \gamma _{eff}=\,$\,  \,$\, \gamma +\gamma _{num}\,\,$\, .

The value of\, \ \,$\, \gamma _{num}\,\,$\, depends on mean time step.
Typically, the latter was between one thirtieth and one twentieth part of
the precession period resulting in \,$\, \gamma _{num}\approx 0.0004\,\,$\, . This
friction is just suitable to simulate good but not best samples. However, it
could be made lower than \,$\, 0.0001\,$\,  if decrease mean time step to about one
fortieth part of the period. Below, the designation \,$\, \gamma \,$\,  will stand for
\,$\, \gamma _{eff}\,\,$\,.

6.6. MSW EXCITATION BY WIRES AND LOOPS.

First of all, excitation of weak magnetic waves in small-area film by wire
and loop inductors was numerically watched for. At present, we confined
ourselves by inductors with round cross-section and radius greater than \,$\, D\,\,$\, %
\ , oriented along \,$\, y\,$\, -axis in parallel to magnetizing field. Corresponding
current induced field, \,$\, h(r,t)=\,$\,  \,$\, \{h_{x}(x,y,t),0,\,$\,  \,$\, h_{z}(x,y,t)\}\,\,$\, , was
calculated from usual magnetostatics formulas. For relations between
physical and dimensionless time and frequency units, see Sec.3.

Examples of such the simulations are illustrated by Figs.6a-c. Two rather
obvious conclusions do follow from these pictures.

(i) At given microwave frequency of linear wire or loop current, not a
single plane wave is induced (as it would be in infinite-size film), but a
spectrum of waves with different length, including ones with non-zero \,$\, y\,$\, %
-component of wave vector (notice that at \,$\, H_{0}=3\,$\,  the
uniform precession frequency \,$\, \omega _{u}\approx \,$\,  \,$\,
6.83\,\,$\, ). In fact, in Fig.6a and Fig.6b we observe eigen-modes %
(or compositions of nearly degenerated eigen-modes) of %
small (finite-size) film. Comparison of %
these figures shows that, naturally, loop inductor ensures better
wave selection and simpler magnetization pattern.

(ii) At the same time, the spatial Fourier spectrum of excited
pattern can be intelligently interpreted in terms of infinite film
theory (Sec.4), as illustrated by contour plot in Fig.6b. In this
plot, a number of lines surrounding some point of \,$\, k\,$\,-plane indicates
its contribution to summary picture. Clearly, spectrum maxima well
agree with equi-frequency curves in \,$\, k\,$\, -plane of infinite film
(Fig.4a in Sec.4), in spite of not large film length to wavelength
ratio (\,$\, \approx 5\,$\,). Hence, so visible characteristic rhombic
structures in Figs.6a-b directly reflect characteristic slopes of
equi-frequency curve responding to the excitation frequency.

The Fig.6c justifies that particular eigen-modes and eigenfrequencies
of even small-size film may be quantitatively close to waves in
infinite film. One particular mode is selected by equating pump
frequency to that of plane surface (Damon-Eshbach) wave with \,$\,
k_{x}=\pm 2\pi /2l\,\,$\, \ and \,$\, k_{y}=0\,$\, which would be
generated by the same loop in infinite film. We see that result is
almost plane wave too, i.e. indeed resonance takes place.
Nevertheless, this mode contains some contribution from plane waves with \,$\, %
k_{x}\approx \pm 3\cdot 2\pi /2l\,\,$\, \ and \,$\, k_{y}\neq 0\,$\,  which possess the
same eigenfrequency and occur resonantly excitable by the same loop.

6.7. ROLE OF\ FILM\ EDGES.

In the top of Fig.7, static distributions of the internal field,
\,$\, W_{0}\,\,$\, , and magnetization in small-area film are shown,
at moderate value of external field, \,$\, H_{0}=3M_{s}\,$\,.
Clearly, in most part of the film practically uniform magnetization
realizes, with \,$\, W_{0}\approx H_{0}\,\,$\, \ . %
Substantial demagnetization takes place at narrow strips only which
adjoin film edges perpendicular to external field and have width
\,$\, \sim D\,\,$\, \ (as it was stated in Sec.3).

Since the internal field is lowered at these demagnetized strips, a local
spin precession frequency there also is lowered. Therefore, usually these
strips take almost none part in shaping and propagation of waves, as if
spins were partially pinned there. These statement can be illustrated by
Figs.6a-c.

In principle, specific edge waves can be excited in the demagnetized
regions. However, this is rather exotic phenomenon, and it was not a case in
practically all of our numerical simulations.

6.8. A SOLITON FED UP BY WEAK CONSTANT\ PARAMETRIC PUMP.

One more exotic phenomenon is illustrated at bottom of Fig.7, concretely,
very small-amplitude soliton (spatially localized wave packet) created and
then supported by extremely weak parametric pump. The loop current parallel
to external field (i.e. to \,$\, y\,$\, -axis) induces field with amplitude of its \,$\, x\,$\, %
-component \,$\, \sim 10^{-4}\,\,$\, (i.e \,$\, \sim 0.01\,\,$\, Oe
in real units) and frequency \,$\, \omega _{e}=14.5\,$\,  which is
far out off total MSW frequency band (with upper bound \,$\,
=H_{0}+2\pi \,$\, \,  \,$\, \approx 9.3\,$\, , see Sec.4). As the
result, long weak envelope soliton is formed whose carrying frequency
equals to half of \,$\, \omega _{e}\,$\,, with magnitude \,$\,
\left\| S_{\bot x}\right\| \sim \,$\,  \,$\, 10^{-5}\,\ \,$\, and
width \,$\, \sim 150D\,$\, \, \ (at film length \,$\, =432D\,\,$\, ).
Interestingly, the two latter quantities approximately satisfy the
relation between amplitude and width of bright solitons which follows
from Eqs.5.15 and 5.25. Although bright solitons moving perpendicular
to external field are formally forbidden in infinite film, this
simulation shows that similar objects are permitted for finite-size
film. Besides, we detect that parametric excitation allows to create
small-amplitude soliton avoiding the restriction (5.23).

Further behavior of this object is even more intriguing. It oscillates
between film ends (see Fig.7) undergoing some decay after reflection from
the distant end but more or less amplification while reflecting from the
edge where inductor takes place. The dot line separates period of strong
amplification of the soliton due to occasionally ``good'' relation between
phases of its carrier and pump. Dimensionless time and frequency in Fig.7
are real ones as expressed in units of \,$\, \,\tau _{0}\,\,$\, and \,$\, f_{0}\,\,$\, ,
respectively (see Sec.2 and 3).

From the point of view of Eq.5.1 the parametric process under discussion
must be described by quadratic term \,$\, \,\left\langle S_{0},h\right\rangle
[S_{0},S_{\bot }]\,\,\ \,$\, in second row. Interestingly, in view of \,$\, %
\,h_{y}=0\, \,$\,  we should conclude that pump, \,$\, \left\langle
S_{0},h\right\rangle \,$\,  , mostly acts at the demagnetized edges. Among our
numerical collection, this is %
exclusive example when film edges play a key role.

Below, we will deal %
with non-parametric pump %
whose frequency belongs to the MSW band. %
Like here, in all forthcoming examples %
inductors are parallel to %
external magnetizing field.

6.9. CREATION OF SOLITONS BY NON-PARAMETRIC PULSE PUMP.

The Fig.8a shows the consequence of intensive radio-impulse of current
passed through wire inductor at one of ends of relatively large-area film.
The impulse duration was about twenty periods of the carry frequency, \,$\, %
\omega _{e}=7.5\,$\,  . At a distance from the inductor, the induced
magnetization precession impulse is deformed. If its initial amplitude was
sufficiently large then further it breaks into a chain of pulses with almost
zero dips between highest of them. The picture in Fig.8a can be qualified as
formation of gray solitons inside finite-length wave packet.

The critical breaking level of amplitude is just its maximum after breaking,
in this example \,$\, \left\| S_{\bot x}\right\| \approx 0.15\,\,$\, \ . This
observation is in reasonable agreement with the estimate of this level which
follows from Eqs.5.9, 5.10 and 5.23,

\begin{equation}
\left\| S_{\bot x}\right\| \approx A_{\min }\sqrt{p}=\sqrt{\Gamma
p/|\varkappa |}\approx 0.1\,
\end{equation}
(at friction \,$\, \gamma =0.0005\,\ \,$\, what took place). In all
the below discussed numerical simulations, close threshold values,
\,$\, 0.1\lesssim \left\| S_{\bot x}\right\| \lesssim 0.2\,\,$\, ,
mark transition to brightly expressed non-linear effects and to
magnetic chaos.

6.10. CHAOS UNDER UNIFORM RESONANT MICROWAVE\ FIELD.

In most of previously reported experiments on magnetic chaos, the latter was
excited by nearly uniform microwave magnetic field, \,$\, h(r,t)\,\,$\, , either using
parallel parametrical pump when \,$\, \,h\Vert S_{0}\,\ \,$\, and excitation frequency
\,$\, \omega _{e}\sim 2\omega _{u}\,\,$\, \ or through perpendicular ferromagnetic
resonance (FMR) when \,$\, \,h\bot S_{0}\,\,\,$\, and \,$\, \omega _{e}\sim \omega _{u}\,\,$\, .
Consider the second variant since it is more close to chaotic
auto-generation of MSW to be under our interest.

The Fig.8b demonstrates modeling of nonlinear FMR in moderate-area film at \,$\, %
H_{0}=3\,\,$\, \ and\, \ \,$\, \gamma =0.0007\,$\, \ , under uniform microwave field
parallel to \,$\, x\,$\, -axis. Due to finite-size effects (edge demagnetization),
factual (numerically found) uniform precession frequency, \,$\, \omega
_{u}\approx 6.785\,$\, \, \ (at \,$\, H_{0}=3\,$\, ), is slightly lower then
theoretical value for infinite film, \,$\, \,\omega _{u\infty }\approx 6.83\,\,$\,.
 Taking \,$\, \omega _{e}\,\,$\, \ sufficiently close to \,$\, \omega
_{u}\,\ \,$\,, it is possible to obtain strong response to weak perpendicular
field \,$\, \,h\sim 0.001M_{s}\sim 0.1\,\,$\, Oe\,. Naturally, the response is indicated
by amplitude of uniform component of \,$\, \,S_{\bot }\,\,$\, \ , i.e. \,$\, \left\langle
S_{\bot }\right\rangle \equiv \int_{V}S_{\bot }dr/V\,\,$\, \ where \,$\, \,V\,\,$\, \
stands for film volume.

After sharp two times increase of pump we observe increase of \,$\,
\left\langle S_{\bot }\right\rangle \,$\, ' amplitude which
monotonically tends to nearly two times larger value. However, next
increases of \,$\, h\,\,$\,  by the same step result in smaller and
non-monotone response. At \,$\, h=0.007\,\,$\, \ , the response
transforms into periodic oscillations. At last, when \,$\,
h=0.008\,\,$\, , these oscillations turn into chaotic one, at time
moment marked as ``burst'' on Fig.8b. More careful repetition of this
process (with smaller step) allows to notice at least one or two
period-doubling bifurcations of regular oscillations (as well know in
theory of dissipative chaos [1]).

Interestingly, to reach the chaos, well satisfied resonance
condition\, \ \,$\, \omega _{e}\approx \omega _{u}\,\,$\, \ is quite necessary.
For instance, at \, \ \,$\, \omega _{e}=\omega _{u\infty }\,\,$\, (i.e. at
frequency deviation less than 1\%\,$\, \,\,$\, ) the only result of even very
intensive pump, \,$\, h\sim 1\,$\, , is strongly nonlinear but regularly oscillating
long-wave structure.

Other characteristic observation is the hysteresis of chaos: if \,$\, h\,$\,  \ is
lowered from \,$\, 0.008\,$\,  than chaotic regime remains at least down to \,$\, h=0.005\,$\,  .

6.11. SHORT-WAVE EXPLOSION AND TRANSITION TO CHAOS.

Usually, experimental magnetic chaos is analyzed in terms of
particular wave modes, that is in momentum space (see, for instance,
[2,3]). In our simulation, we can view also how it looks in real
space, and watch for spatial-temporal picture of transition from
regular motion to chaos. Characteristic scenario of this transition
is illustrated in Fig.9.

When FMR is still regular, rather smooth magnetization pattern takes place
with one or three maximums of oscillations and precise mirror symmetry (with
respect to middle lines of rectangular film area). However, the closer is
the transition the higher and narrower is the central maximum. This means
that spectrum of excited MSW becomes more and more wide, but still coherent,
in the sense that all the waves are mutually connected by some rigid phase
relations. At critical ``burst'' time moment the central maximium collapses
into peak very narrow in \,$\, x\,$-direction and rather flat (elonged) in \,$\, y\,$\, %
-direction (i.e. along static magnetization). Then this peak blows up
giving freedom to to the short waves. The latter incoherently scatter
in all the directions in their turn giving rise to complicated
(chaotic) magnetization pattern. Beginning of this unstable explosive
stage is shown at bottom of Fiq.9. Notice that corresponding
magnitude of \,$\, S_{\bot x}\,\,\,$\, confirms estimate (2).

Important sign of transition to chaos is violation of the mirror symmetry.
The symmetry with respect to \,$\, 180^{o}\,\,$\, rotation only remains after the
explosion. Clearly, this symmetry is invoked by that of the static \,$\, S_{0}\,\,$\, %
\ pattern (see Fig.7).

At later time, most short of the explosively induced waves decay. In further
stationary chaotic regime, magnetization picture more or less restores both
smoothness and mirror symmetry. But, naturally, increase of pump rises both
short-wave contents and asymmetry. Moreover, under sufficiently intensive
pump new similar explosions (bursts) are repeated from time to time, serving
as ``discharges'' of excessive energy accumulated by long waves. This may be
called strong chaos.

6.12. EXCESS ENERGY\ AND POWER\ ABSORPTION.

The top of Fig.9 demonstrates typical chaotic behavior of excess film
energy, \,$\, E\,$\,  , and of power absorption by film, \,$\,
P\,\,$\, , both related to unit volume (discretization cell). Here
and below, ``excess energy'' (or simply
``energy'') will term increase of magnetic energy due to the excitation, \,$\, %
S_{\bot }\,\,\,$\, , but excluding direct \,$\, \,S_{\bot }\,$\, 's interaction\,$\, \,\,$\, \ with
pump field \ (i.e. except \,$\, \,-\int \left\langle h,S_{\bot }\right\rangle
dr\,/V\,\,$\, \ , see Sec.2). The power absorption, \,$\, P\,$\,  , describes energy flow
into film which is spent for both \,$\, E\,\,$\, \ and dissipation in the film
interior. Hence, in general

\begin{equation}
\frac{dE}{dt}=P-P_{dis}\,\ ,\,\,\ \ \ \,P=\left\langle \int \left\langle
h(r,t),\frac{dS_{\bot }}{dt}\right\rangle \frac{dr}{V}\right\rangle _{T}\,\ ,
\end{equation}
with \,$\, P_{dis}\,\,$\, \ being dissipated power per unit volume. Symbol\, \ \,$\, %
\,\left\langle {}\right\rangle _{T}\,$\,  \, \ designates time averaging
with respect to spin precession. These equations directly follow from the
basic Eq.2.1.

It is natural to expect that

\begin{equation}
P_{dis}\approx 2\Gamma (E-E_{0})\,\,\,,
\end{equation}
at some \,$\, E_{0}\,$\,  , and \,$\, \Gamma \,\,$\, \ being previously considered dissipation
rate of magnetization. At \ \,$\, \gamma =0.0007\,$\, \ first of Eqs.5.10 gives \,$\, %
2\Gamma \approx 0.013\,\,$\, , then energy relaxation time is\,$\, \,\sim 1/2\Gamma
\approx 80\,\,$\, . Indeed, approximately such the time scale can be viewed at
plot (A) in Fig.9. Further, let us put on \,$\, \,\left\langle S_{\bot
x}\right\rangle =\,$\,  \,$\, \left\| S_{x}\right\| \sin [\omega _{e}t-\phi ]\,\,\,$\,
where \,$\, \left\| S_{x}\right\| \,$\,  and \,$\, \phi \,\,$\,  designate amplitude and phase,
respectively, of \ the uniform component of \ spin precession. Then

\begin{equation}
P=h\left\langle \sin (\omega _{e}t)\frac{d\left\langle S_{\bot
x}\right\rangle }{dt}\right\rangle _{T}\approx \frac{1}{2}\,\omega\,
_{e}h\left\| S_{x}\right\| \sin \,\phi \,
\end{equation}
At \,$\, \left\| S_{x}\right\| \approx 0.1\,\,$\, (as prompted by Fig.8b), \,$\, h=0.008\,\,$\, %
\ and \,$\, \phi \approx \pi /2\,\ \,$\, (which means good resonance) this relation
results in \,$\, P\sim 2.5\cdot 10^{-3}\,\,$\, , in agreement with plot (B) in Fig.9.
Hence, rough average characteristics of chaotic variables are easy
explainable.

Much harder task is to explain chaotic deviations from average
values. For instance, in plots (A) and (B) energy and power nearly
follow one another. It is clear: the energy comes from power
absorption, but the latter depends on the phase \,$\, \phi \,\,$\, \
which in its turn must be sensitive to energy, because of
non-isochronity of spin precession (see Sec.5). But without an
adequate dynamical model for this connection we can not estimate
details of chaotic time series.

6.13. INSTANT FREQUENCY.

Directly, our numerical algorithm produces fast oscillating time series. To
extract from them relatively slow \ time-varying (``instant'') amplitudes
and phases (like \,$\, \left\| S_{x}\right\| \,$\,  and \,$\, \phi \,$\,  above), and
corresponding instant frequencies (e.g. \,$\, \omega _{e}-d\phi /dt\,\,$\, \ ), there
are two ways. One is to build analytical signals by means of (discrete)
Gilbert transformation. Another way is to indicate maximums, minimums and
zero-crossings of oscillating variable. If \,$\, \,t_{n}\,\,$\,\,, \, %
\,$\, n=\,...,-1,0,1,...\,$\,, are estimates of time moments when zero-crossing does
occur, then instant frequency at \,$\, t\approx t_{n}\,\,$\, \ can be determined as \,$\, %
\omega _{in}=\,$\,  \,$\, 2k\pi /(t_{n+k}-t_{n-k})\,\,$\, \ , where \,$\, k\geq 1\,\,$\, .

If disposing this quantity,the phase can be restored by discrete numerical
integration. This method is more comfortable and fast than Hilbert
transformation, but ensures not worse (usually better) accuracy what was
confirmed by special tests.

6.14. QUASI-LOCAL ENERGY DENSITY.

Because of relations (2.9) and (4.1), the excess energy (per unit volume)
can be represented in the form

\begin{equation}
E=E_{loc}+E_{nonloc}\,\ ,\,\ \,\ \ \ E_{loc}=\int e_{loc}\,\frac{dr}{V}\
\,,\,\ \ \,\ e_{loc}\equiv H_{0}(1-S_{y})+2\pi S_{z}^{2}\,
\end{equation}
Here \,  \,$\, E_{nonloc}\,\,$\, \ is contribution from non-local
(non-singular) part of dipole interaction in plate geometry (Sec.4), while \,$\, %
E_{loc}\,\,$\, consists of its local (singular) part and also local first term of
Eq.2.9.

For uniform precession (even let large-amplitude and strongly non-linear),
\,  \,$\, E_{nonloc}\,\,\,$\, vanishes, hence, the sum \,$\, \,e_{loc}-\left\langle
h(t),S\right\rangle \,\,$\,  plays the role of Hamiltonian of the average
spin. In particular, in autonomous regime of uniform precession, at \,$\, h=0\,$\,  , \,$\, %
\,\,e_{loc}\,\,$\, \ becomes integral of motion.

Importantly, \,$\, E_{nonloc}\,\,$\,  may be negligible as compared with \,$\, E_{loc}\,\,$\, in
non-uniform chaotic case too. Indeed, according to equations (4.1-2), (4.6)
and (6), the non-local contribution can be estimated as

\begin{equation}
E_{nonloc}\approx \pi D\int \left( \frac{k_{x}^{2}}{|k|}\left| \widetilde{S}%
_{x}(k)\right| ^{2}-|k|\left| \widetilde{S}_{z}(k)\right| ^{2}\right) dk\,\
,\,\
\end{equation}

\[
\left| \frac{E_{nonloc}}{E_{loc}}\right| \ \lesssim \frac{\pi D\left\langle
|k|\right\rangle }{H_{0}}\,\ ,\,\ \ \,\left\langle |k|\right\rangle \equiv
\int |k|\left| \widetilde{S}_{x}(k)\right| ^{2}dk\left( \int \left|
\widetilde{S}_{x}(k)\right| ^{2}dk\right) ^{-1}\,,
\]
with \,$\, \widetilde{S}_{x,z}(k)\,$\,  \,$\, \,\,$\, denoting spatial Fourier transform of
magnetization. If long MSW (with \,$\, D|k|\ll 1\,$\,  ) are dominating in
magnetization pattern, then the excess energy is well characterizable by \,$\, %
E_{loc}\,\,$\, . As the consequence, the quantity \,$\, \,e_{loc}\,\,$\,  behaves like
local integral of motion and thus can be termed quasi-local energy density.

6.15. WEAKLY CHAOTIC FMR.

Seemingly weak chaos realizes at \,$\, h=0.005\,\,$\, . The Fig.10 shows evidence for
(i) good correlation between the energy and dissipation (with \,$\, E_{0}=0\,\,$\, \
and \,$\, \Gamma \,\,$\, \ well related to actual friction coefficient \,$\, \gamma \,\,$\, ),
and, at plot (D), (ii) rigid correlation between energy and average
longitudinal component of magnetization, \,$\, \left\langle S_{\Vert
}\right\rangle \,\,$\, . Here \,$\, \,\left\langle ..\right\rangle \,\,$\, \ denotes
space-time averaging. Notice that in most part of film area \,$\, S_{\Vert }=\,$\,  \,$\, %
\left\langle S_{0},S\right\rangle \approx \,$\,  \,$\, S_{y}\,$\,.

The latter correlation just gives the evidence that non-local energy
contribution is relatively small. Indeed, because of relations

\[
\left\| S_{z}\right\| ^{2}\approx \left\| S_{x}\right\| ^{2}/p^{2}\,\ ,\,\ \
\,\,\left\langle S_{x}^{2}\right\rangle +\left\langle S_{z}^{2}\right\rangle
\approx 2(1-\left\langle S_{\Vert }\right\rangle )\,,
\]
where \,$\, p\,\,$\, \ is characteristic eccentricity, \,$\, \,E_{loc}\,\,$\, \ can be expressed
as\

\begin{equation}
E_{loc}\approx \left( H_{0}+\frac{4\pi }{p^{2}+1}\right) (1-\left\langle
S_{\Vert }\right\rangle )
\end{equation}
Hence, at \,$\, E_{nonloc}\,$\,  \,$\, \ll E_{loc}\,\,$\,  total excess energy also well reduces
to \,$\, \left\langle S_{\Vert }\right\rangle \,\,$\,. If we equated \,$\, p\,\,$\, \ to
small-amplitude uniform precession eccentricity given by Eq.5.10, then at \,$\, %
H_{0}=3\,$\,  the Eq.8 would yield \ \,$\, dE/d\left\langle S_{\Vert }\right\rangle
\approx \,$\,  \,$\, dE_{loc}/d\left\langle S_{\Vert }\right\rangle \approx \,$\,  \,$\, -5\,\,$\, .
This value is lower by 10\% than the slope (\,$\, \approx -5.6\,$\, ) at Fig.10.D. The
difference can be explained if take into account that increase in spatial
non-uniformity of precession result in decrease of its eccentricity.

Plot (G) in Fig.10 presents spectrum of power absorption (absolute value of
Fourier transform of \,$\, \ P(t)\,\,$\, ). At \,$\, M_{s}\approx 140\,\,$\, \,
Oe\,  (as for YIG), the dimensionless frequency unit corresponds to \,$\, %
\approx 390\,\,$\, \,  MHz, hence, the dominating frequency in \,$\,
P(t)\,$\, 's spectrum is \,$\, \approx 20\,\,\,$\, MHz. Chaos in
instant frequency of precession, as shown in plot (B), and thus in
its phase is characterized by significantly wider frequency band (in
part contained by plot (C) in Fig.9). Nevertheless, plot (F) in Fig10
visualizes anti-correlation between instant frequency and energy
which corresponds to negative sign of non-isochronity (see Sec.5).

For fractal dimension (Sec.5.9) of these data it was found that \,$\, %
\,d_{cor}<3\,\ \,$\, (see below). This implies that chaos is governed by three
relevant variables only, and therefore chaotic attractor could be
represented in 3-dimensional space. It seems doubtless that \,$\, \,E\,\,$\, \ , \,$\, P\,$\,  \
and \,$\, \left\| S_{x}\right\| \,$\,  are relevant variables (due to Eq.5, \,$\, P\,$\,  \ and \,$\, %
\left\| S_{x}\right\| \,$\,  determine also the phase\,  \,$\, \,\phi \,\,$\, ).
Plot (C) in Fig.10 gives show of the attractor in these coordinates.

6.16. ANALYSIS OF\ CORRELATION\ DIMENSION.

Let us discuss practical calculation of the correlation dimension, \,$\, %
\,d_{cor} \,$\,  (Sec.5.9). At given correlation sum, one may estimate \,$\, \,d_{cor}\,$\,
by two ways:

\begin{equation}
d_{cor}=\frac{d\ln \,\sigma (R)}{d\ln \,R}\,\,\ \ \ \text{or\,  }\,\ \
\ d_{cor}=\frac{\ln [\,\sigma (R)/\Omega ]}{\ln \,R}\
\end{equation}
The first of them may be called differential dimension while the
second integral. Under formal limit \,$\, N\rightarrow \infty \,\,$\,
\, \ these quantities are expected to coincide one with another. But
real calculation needs in more than \,$\, N^{2}\,$\,  operations,
therefore \,$\, N\,$\,  can not be as large as wanted. Under
realistic \,$\, N\,\,$\, \ , the differential estimate satisfactorily
works at moderate values of \,$\, R\,\,$\, \ only, \,$\, \,R_{\min
}\ll \,$\,  \,$\, R\ll \,$\,  \,$\, R_{\max }\,\ \,$\, (where \,$\,
R_{\min }\,$\,  is minimum of \,$\, R_{ij}\,\,$\, )\,$\, .\,$\,  In
opposite, the integral estimate better works at lower end of this
interval.

Naturally, the integral estimate is less sensitive to finiteness of
\,$\, N\,$\,  , but instead it requires to know the coefficient \,$\,
\,\Omega \,$\,  . It is rather obvious that the best general
assumption about \,$\, \,\Omega \,$\,  is that it equals to volume of
\,$\, \,d_{cor}\,$\, -dimensional unit-radius sphere, i.e.\,$\,
\;\Omega =\pi ^{\,d_{cor}/2}/\Gamma (1+\,$\,  \,$\,
\,d_{cor}/2)\,\,$\, \,  , \ where \,$\, \Gamma \,\,$\, \ means
gamma-function. By special tests we verified that this recipe indeed
constantly improves precision of \,$\, \,d_{cor}\,$\,  estimates.

Notice that wide class of tests is presented by the Kaplan-Yorke chaotic
system [1]. It is described by the set of difference (discrete time)
equations:

\begin{equation}
X_{j}(t+1)=F(X_{j}(t))\,\ ,\,\ j=1 ...n\,\,,\,\ \ Y(t+1)=\alpha
Y(t)+\sum_{j=1}^{n}f_{j}(X_{j}(t))\,\ ,\,\,\ \ |\alpha |<1\,\ ,
\end{equation}
where \,$\, F(X)\,$\,  is some chaotic one-dimensional map (for instance, tent map, \,$\, %
F(X)=1-|2X-1|\,\,\,$\, ), and \,$\, \,f_{j}(X)\,$\,  are any smooth functions. Dependently
on \,$\, \,n\,\,$\, \ and \,$\, \,\alpha \,$\,  , fractal dimension of chaotic sequence \,$\, Y(t)\,$\, %
\, \ can be equated to arbitrary number (for example, for the
tent map \,$\, n=2\,$\,  and \,$\, \,\alpha =1/16\,$\,  lead to \,$\, \,d_{cor}=2.5\,$\,  while
\,$\, n=3 \,$\,  and \,$\, \alpha =1/64\,\,$\,  to \,$\, \,d_{cor}=3.5\,$).

6.17. FRACTAL DIMENSION OF\ CHAOTIC\ FMR.

Typical example of evaluation of \,$\, \,d_{cor}\,$\,  is shown in Fiq.11a (left
plot). Here ``log of Cell Size'' in horizontal axis means \,$\, \,\ln \,(R_{\max
}/R)\,\,$\, \ while vertical axis presents both the estimates (9). The
dependencies of differential and integral dimensions on \,$\, \,\ln \,(R_{\max
}/R)\,\,$\, \ are drawn by thin and fat lines, respectively.

We see that plateau in the first of them well coincides with the upper value
of the second. From above mentioned tests, it is known that just this value
should be taken as best estimate of \,$\, \,d_{cor}\,$\,  , and that the coincidence
signify good reliability of this estimate. In more complicated case when the
plateau differs from upper (most right-hand) value of integral dimension,
the latter must be preferred. But such the situation testifies that either
the data (finite chaotic series) are not enough representative or they
possess essential multi-fractality.

In case under consideration, fractal dimension of the power absorption time
series under weakly chaotic FMR can be estimated as \,$\, \,d_{cor}\approx 2.4\,\,$\,.

To form better representative data, the time separation, \,$\, \,\tau \,$\,  , in \,$\, d\,$\, %
-dimensional embedding points \,$\, \{x(t_{0}+n\tau ),\,$\,  \,$\, x(t_{0}+n\tau +\tau ),\,$\,  \,$\, %
...,x(t_{0}+n\tau +d\tau )\}\,$\,  should be a few times shorter than
characteristic correlation time of \,$\, \,x(t)\,\,$\, , while number of points must
be sufficient for minimum ``filling'' of all of \,$\, \,d\,$\,  dimensions, at least \,$\, %
N\geq 2^{d}\,\,$\,.

6.18. CHAOTIC FMR IN DEFECTIVE\ FILM.

Real films always have more or less amount of defects. The right-hand plot
in Fig.11a demonstrates static magnetization by tangential field in film
with periodic lattice of defects (punctures) which touch about 10\% of film
area. Naturally, mean internal field, \,$\, W_{0}\,\,$\, , and thus characteristic
precession frequencies are lowered by defects. But we found no qualitative
difference between chaotic FMR in defective film and ``good'' film.

The top and left bottom plots in Fig.11b illustrate how the uniform
component of spin precession behaves at small time scale (of order of \,$\, %
\,1\,ns\,$\,) and at moderate time scale (of order of \,$\, \,0.1\,\mu s\,\,$\,). On
right hand, contour plot of spatial Fourier transform of \,$\, \,S_{\bot x}\,\,$\,
shows that two groups of MSW modes form magnetization pattern, both
belonging to the same frequency range around \,$\, \omega _{u}\,$\,  , with the
short-wave group induced by defect lattice. The levels for this plot were
chosen specially to highlight short-wave modes. In fact, their contribution
to energy is of order of a few percents.

6.19. SYNCHRONIZATION\ OF\ CHAOTIC\ FMR.

In the work [4] synchronization of magnetic chaos under non-linear FMR in
normally magnetized YIG film was experimentally realized. First, the signal, \,$\, %
P_{M}(t)\,$\,  , related to power absorption was recorded into a memory.
Characteristic frequencies of this signal were between 0.5 MHz and 10 MHz.
The similar actual ``slave''signal, \,$\, P_{S}(t)\,$\,  , was compared with the
recorded ``master'' signal, and the difference was directed, with some
proportionality constant, \,$\, K\,$\,  , to perturb the external magnetizing field, \,$\, %
\,H_{0}\rightarrow \,$\,  \,$\, H_{0}+K(\,$\,  \,$\, P_{M}-P_{S})\,$\,  . At suitable choice of \,$\, K\,\,$\, %
, after a transient time \,$\, \,\sim 10\,\mu $s\,, excellent coincidence between \,$\, %
P_{S}(t)\,$\,  and \,$\, P_{M}(t)\,$\,  was observed.

The Fig.12 illustrates the attempt to numerically reproduce such the
experiment but with tangentially magnetized film. The above mentioned
defective film model is explored, at \,$\, H_{0}=3\,\,$\, \ and \,$\, \,h=0.005\,$\, . The power
absorption per unit volume, \,$\, P(t)\,$\,  , is taken to serve as the control
signal. The master signal to be addressed to the feedback, \,$\, P_{M}(t)\,$\,  ,
either equals to \,$\, P(t)\,$\,  or formed from it by slight time-smoothing (over \,$\, %
5\div 30\,$\,  periods of precession). Typical magnitude of chaotic \,$\, P(t)\,$\, 's
variations is \,$\, \sim 5\cdot 10^{-5}\,\,$\, . The feedback coefficient, \,$\, K\,\,$\, , is
changed in the range between \,$\, \,-30\,\,$\, \ and \,$\, -600\,\,$\,.

Unfortunately, \,$\, 10\,\mu $s\, is rather large time for our numerical
simulations. Total duration of numerical runs was just about \,$\, 10\,\mu $s\,.
Best signs of synchronization were observed at \,$\, K\sim 200\,\,$\, . Hence,
magnitude of the bias field modulation was \,$\, \sim 0.01\,\,$\,, %
i.e. \,$\, \sim 1.5\,\,$ Oe in real units. %
For comparison, in [4] essentially smaller values
\,$\, \sim 0.1\,\,$\, Oe were in action.

According to Fig.12, with no doubts synchronization takes place, but its
quality is far from so excellent as reported in [4]. Possibly, this is due
to wider frequency range of the control signal in our system, up to \,$\, \sim
50\,\,$\, \ MHz (see plot of \,$\, P(t)\,$'s spectra in Fig.12) %
and to not long enough duration of the numeric experiment.

\,\,

{\it REFERENCES}

1. A.J.Lichtenberg and M.A.Lieberman. Regular and stochastic motion.
Springer-Verlag, 1988.

2. S.M.Rezende and F.M.de Aguiar. Proc. IEEE, 78 (1990) 893.

3. J.Beeker, F.Rodelsperger,Th.Weyrauch, H.Benner, W.Just and
A.Cenys. Phys.Rev. E59 (1999) 1622.

4. D.W.Peterman, M.Ye and P.E.Wigen. Phys.Rev.Lett. 74 (1995) 1740.

\,\,

\section*{Conclusion}

The above expounded material leads to %
conclusions as follow: %

(i) Concepts, formulas and observations of linear theory of %
magnetostatic spin waves (MSW) in infinite films %
(Sections 2-4 of Part I, see\, arXiv preprint\, 1204.0200\,), %
as well as that of quasi-linear MSW theory (Sec.5 in the Part I), %
appear quite adequately useful for interpretation of numeric %
simulations of even non-linear and chaotic MSW %
even in small-size films;\,

(ii)  Most effective way to MSW chaos is via %
parametric resonance and parametric non-linear transformations of
MSW;\,

(iii) MSW chaos typically is determined by a few relevant %
variables only, - i.e. characterized by rather low fractal %
dimension , - and therefore seems allowing its control %
and more or less satisfactory synchronization.

\,\,\, 

The two last features will be further investigated %
in next Part III of this manuscript, %
concentrating on not externally driven %
but auto-generated MSW chaos.



\newpage  \topmargin -2 cm

\begin{figure}
\includegraphics{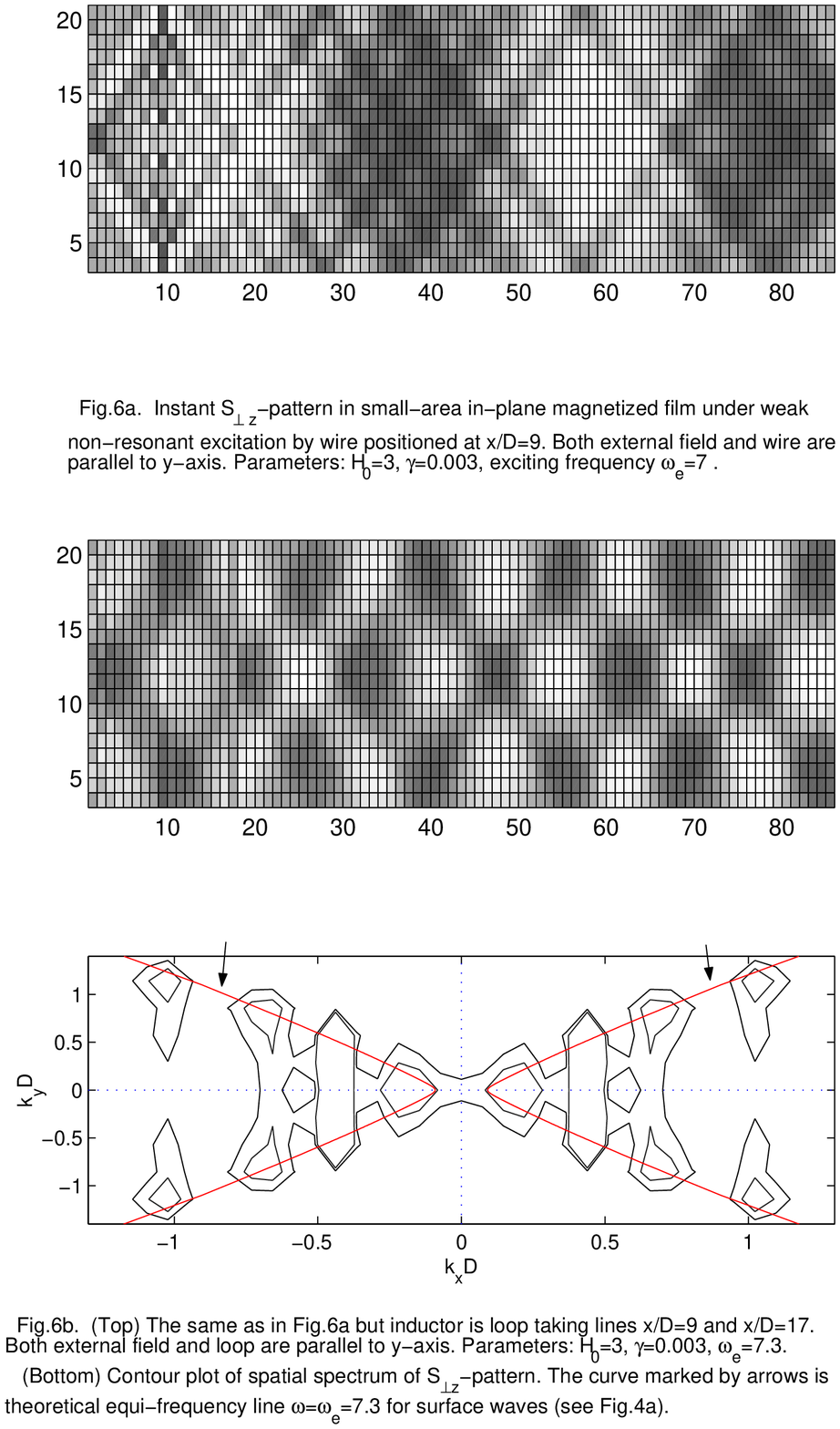}
\end{figure}

\newpage
\begin{figure}
\includegraphics{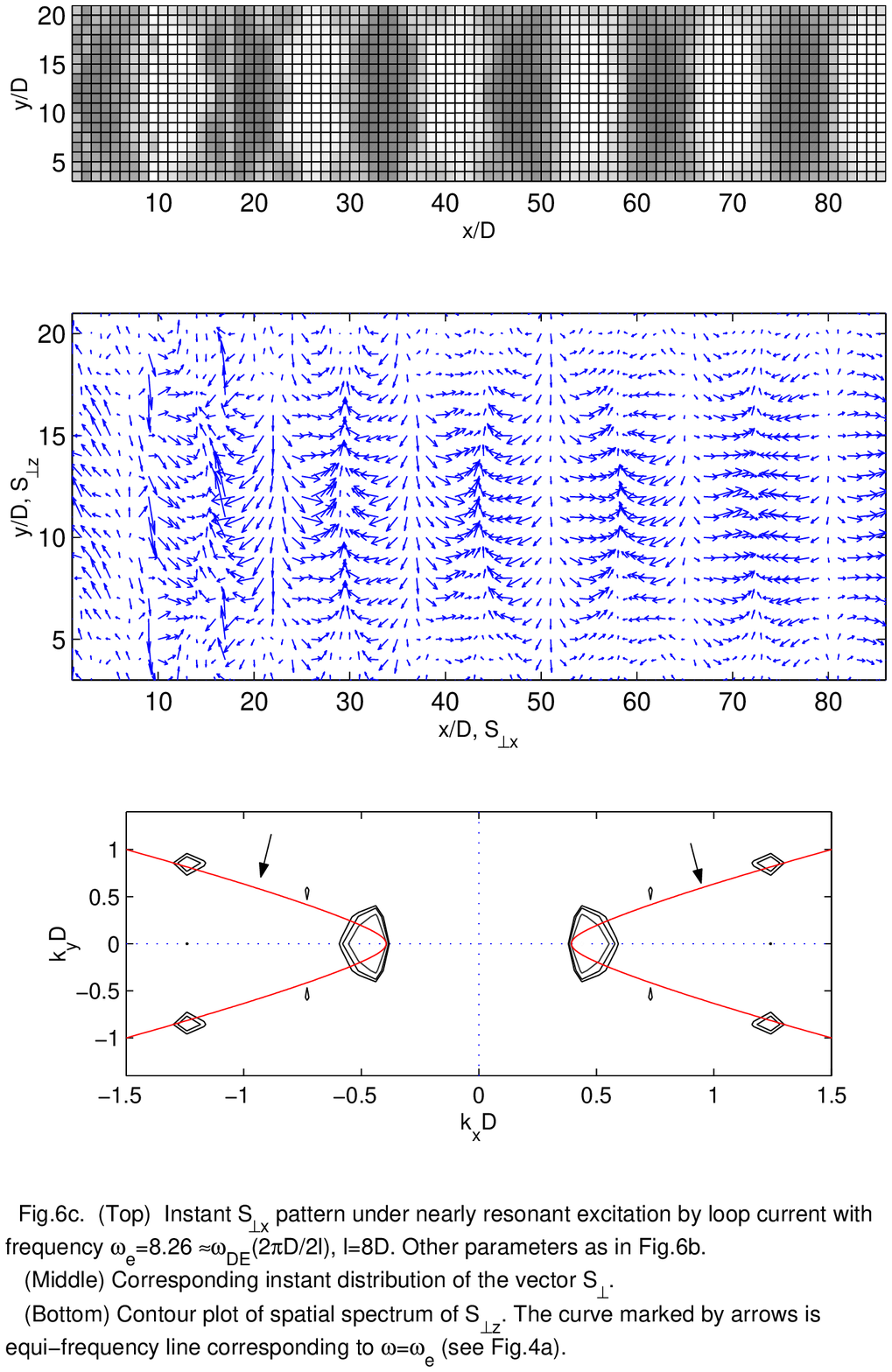}
\end{figure}

\newpage

\begin{figure}
\includegraphics{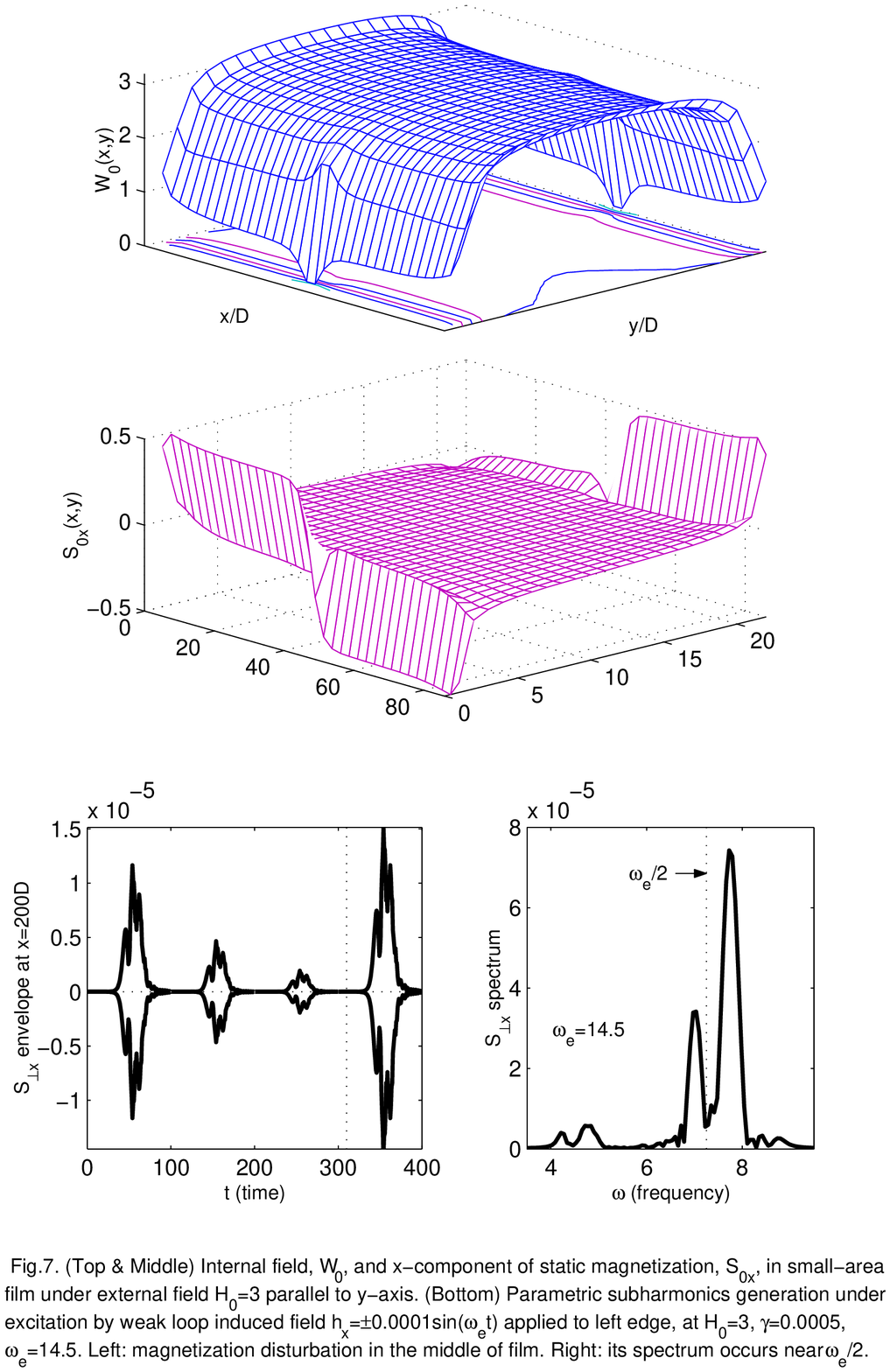}
\end{figure}

\newpage

\begin{figure}
\includegraphics{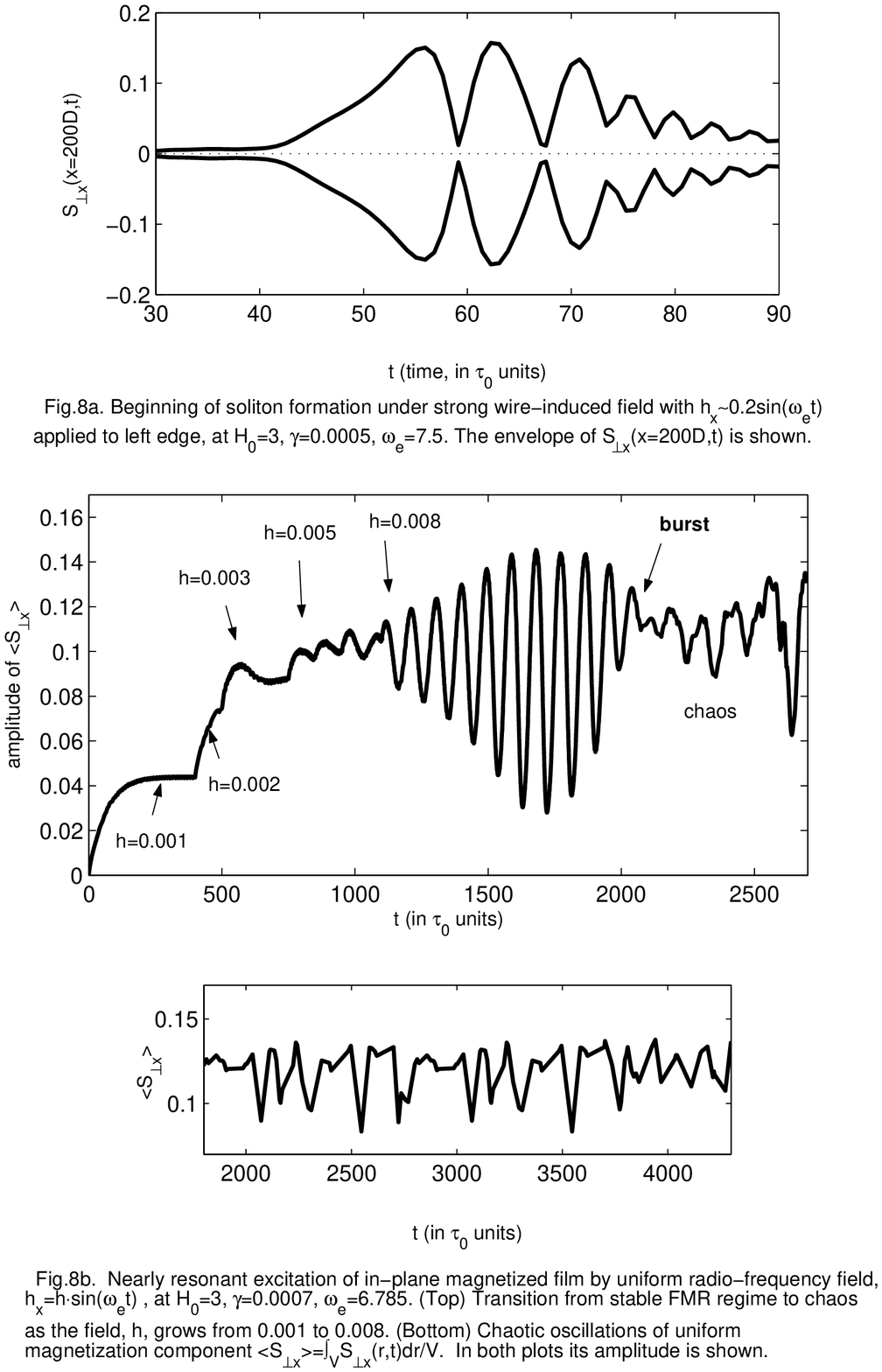}
\end{figure}

\newpage

\begin{figure}
\includegraphics{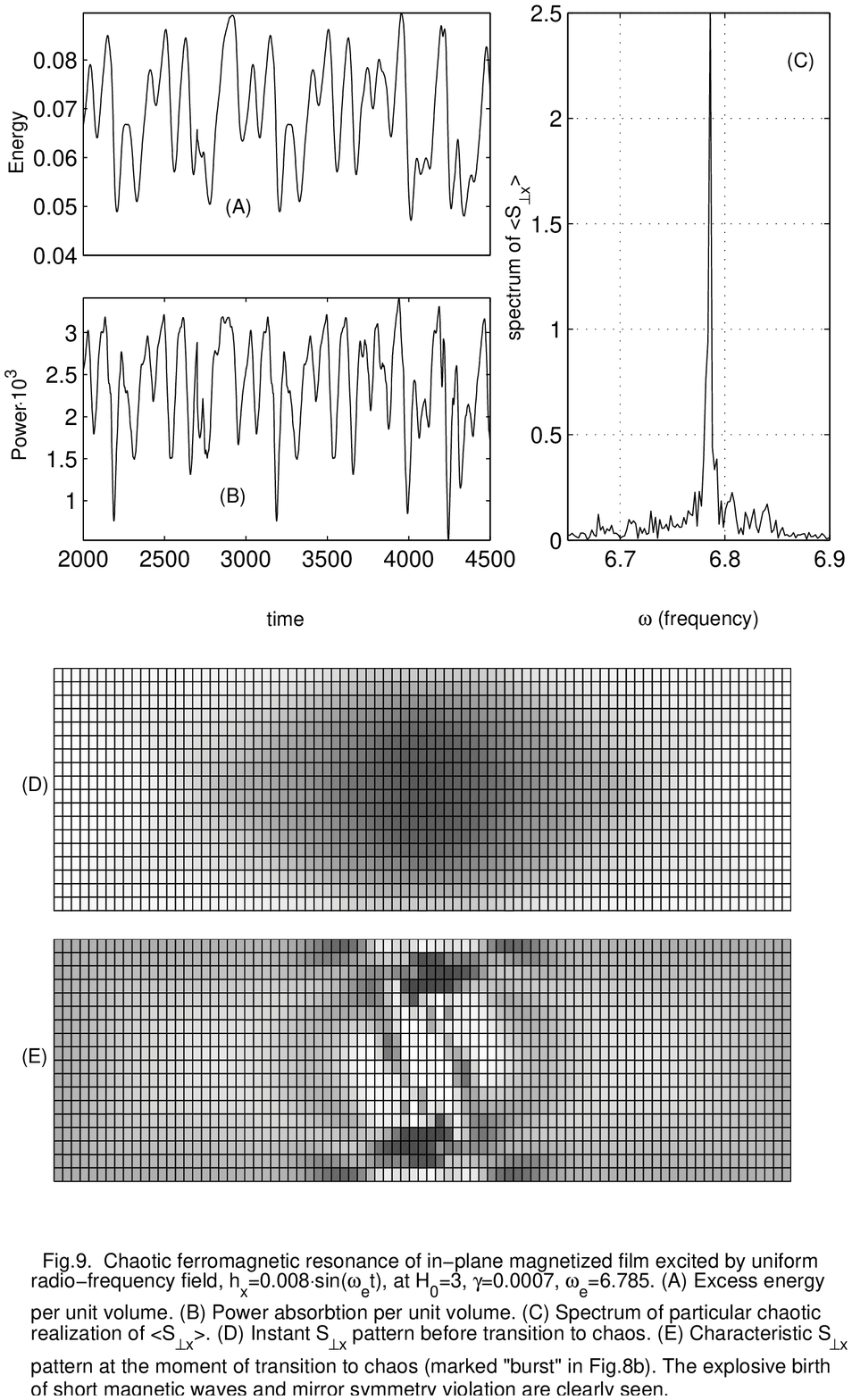}
\end{figure}

\newpage

\begin{figure}
\includegraphics{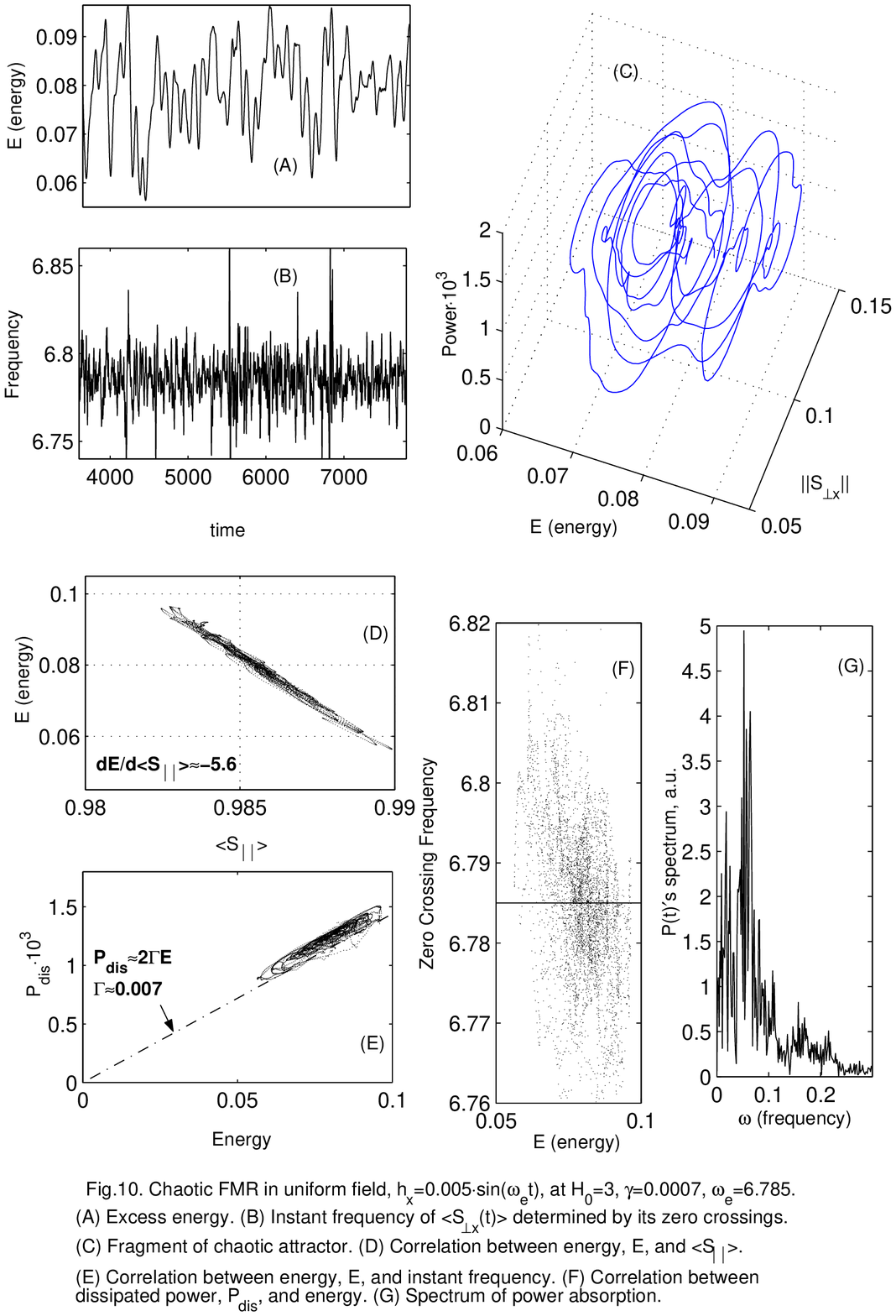}
\end{figure}

\newpage

\begin{figure}
\includegraphics{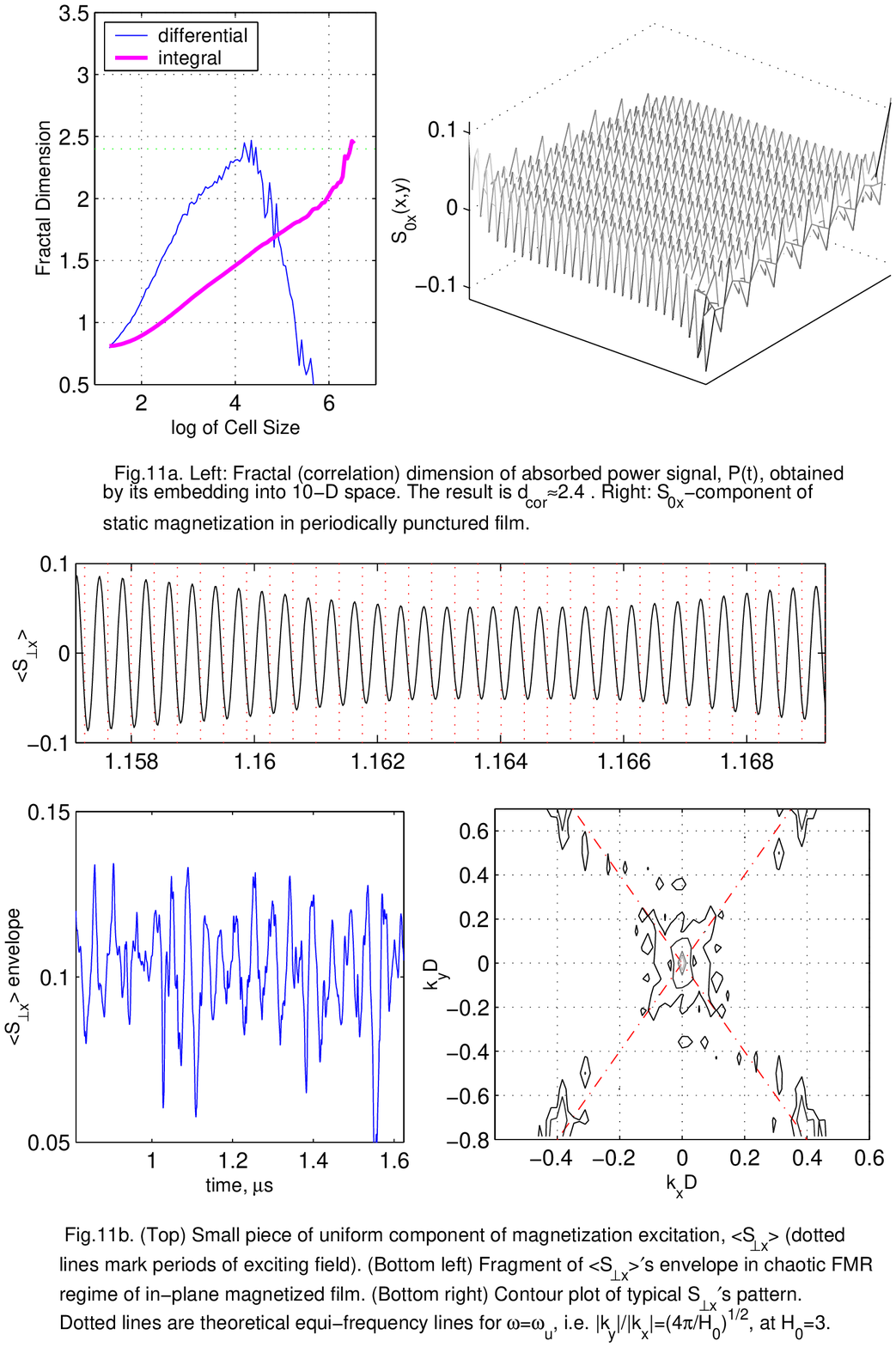}
\end{figure}

\newpage

\begin{figure}
\includegraphics{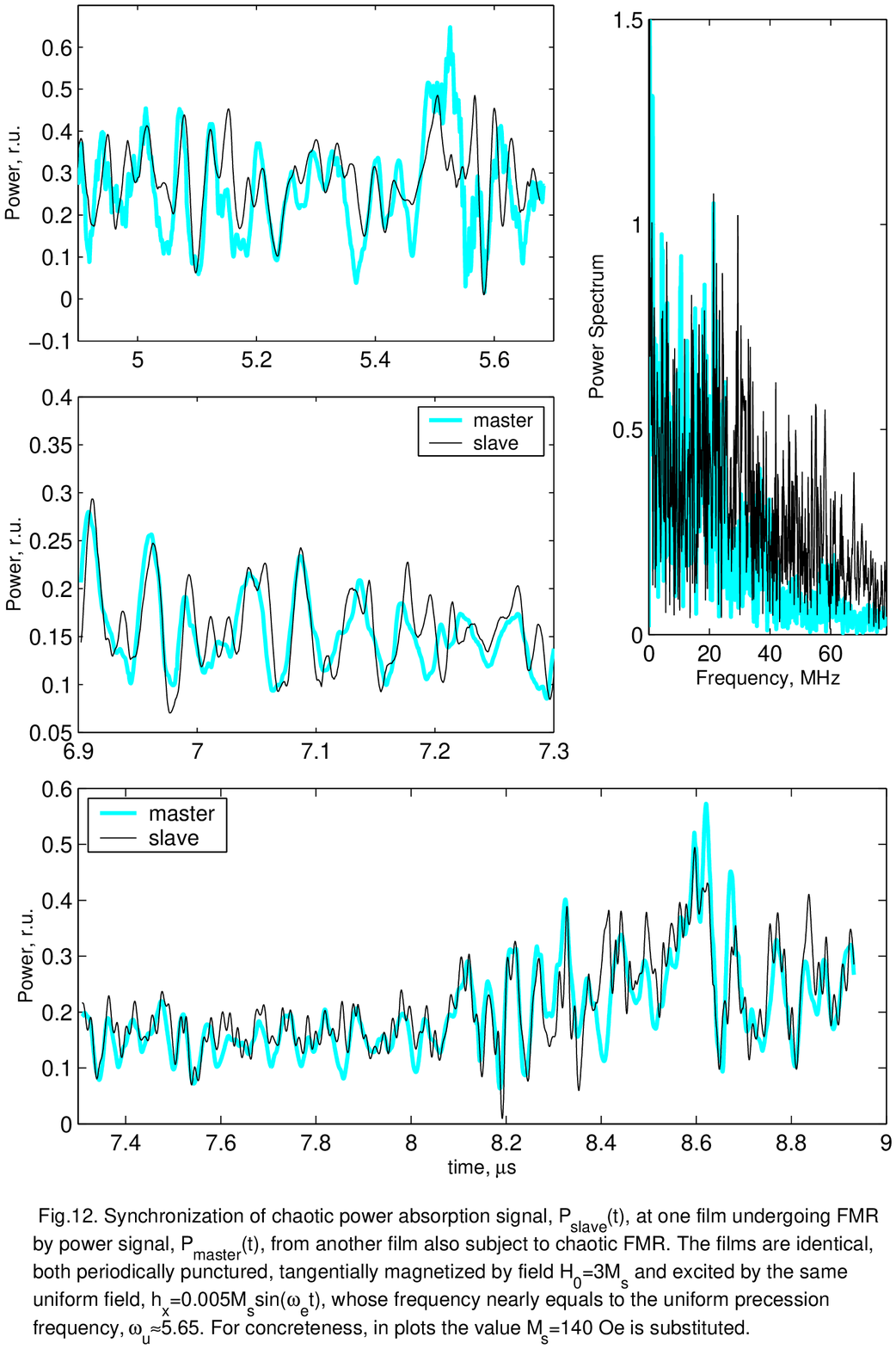}
\end{figure}




\end{document}